\documentclass[11pt]{article}

\usepackage[T1]{fontenc}
\usepackage[margin=1in]{geometry}
\usepackage{amsmath,amssymb,amsthm}
\usepackage{graphicx}
\usepackage{xcolor}
\usepackage{hyperref}
\usepackage{bbm}

\hypersetup{
    colorlinks=true,
    linkcolor={black!70!blue},
    citecolor={black!70!blue},
    urlcolor={black!70!blue}
}

\title{\textbf{Stochastic Dynamics of Large-Scale Motif-Embedded Spiking Neuronal Networks}}
\author{
Gurpreet Jagdev$^{1,2}$, Richard Bertram$^{3,4,5}$, Na Yu$^{1,2}$ \\
{\small $^{1}$Department of Mathematics, Toronto Metropolitan University, Toronto, Canada}\\
{\small $^{2}$Institute for Biomedical Engineering, Science and Technology (iBEST), Toronto Metropolitan}\\ {\small University and St. Michael's Hospital, Unity Health Toronto, Toronto, Canada}\\
{\small $^{3}$Department of Mathematics, Florida State University, Tallahassee, Florida, USA}\\
{\small $^{4}$Program in Neuroscience, Florida State University, Tallahassee, Florida, USA}\\
{\small $^{5}$Institute of Molecular Biophysics, Florida State University, Tallahassee, Florida, USA}
}
\date{}

\begin{document}
\maketitle

\begin{abstract}
\label{abstract}
We examine how local motif structure and global network topology jointly shape spiking dynamics in stochastic neuronal networks. Using networks of Izhikevich neurons with Erd\H{o}s-R\'enyi (ER) and scale-free (SF) background connectivity, we compare motif-embedded networks with synapse-count-matched, non-motif controls under noise- and stimulus-driven protocols. Motif embedding increases noise-induced coherence in both topologies and provides a smaller improvement in signal transmission, while SF-based networks attain greater absolute coherence and transmit signals faster and more reliably than ER-based networks. Motif-rewiring experiments show that this benefit depends on the specific arrangement of intra-motif connections rather than on strong local connectivity alone: randomizing the motif skeleton reduces coherence overall, although rewiring the non-recurrent feed-forward loop and bi-parallel motif can increase it. Targeted ablation shows that the SF advantage depends in part on hub integrity; removing high-degree excitatory neurons impairs coherence and signal transmission more than matched random node or edge removal. Together, these results indicate that local motifs and global hubs make distinct, complementary contributions to coherent spiking dynamics.
\end{abstract}

\vspace{0.5em}

\section{Introduction}
\label{sec:introduction}
Understanding how network structure shapes neural dynamics is a central question in computational neuroscience~\cite{bullmore2009complex}. Population activity depends not only on the intrinsic dynamics of individual neurons, but also on the organization of synaptic connections among them. Network structure therefore influences whether spiking activity remains local and dissipates or propagates into coherent population activity~\cite{van2002fast,pisarchik2023coherence}.

Noise also shapes neuronal dynamics. Biological systems operate amid persistent fluctuations---including the quasi-random release of neurotransmitters, stochastic synaptic transmission, and sensory or environmental noise~\cite{Nolte2019CorticalReliability,faisal2008noise}. Such randomness can degrade temporal precision, but may also enhance the coherence of nonlinear excitable systems. Coherence resonance (CR)~\cite{gang1993stochastic,pikovsky1997coherence} and related forms of stochastic facilitation~\cite{mcdonnell2011benefits,forgoston2018primer} exemplify this dual role: in a system without periodic forcing, an intermediate noise intensity can produce the most coherent response. In a network setting, CR therefore depends not only on the stochastic forcing but also on the structure through which noise-driven activity spreads~\cite{pisarchik2023coherence}.

At the global scale, Erd\H{o}s-R\'enyi (ER) networks provide a comparatively homogeneous model of random connectivity, with degree concentrated near its mean and no systematic class of high-degree hubs~\cite{albert2002statistical}. Scale-free (SF) networks, by contrast, have heavy-tailed degree distributions, with a small number of highly connected hubs among many sparsely connected nodes~\cite{barabasi1999emergence}. Such hubs can shorten paths and distribute activity across the network~\cite{van2011rich}, and previous work suggests that heterogeneous connectivity can support coherent spiking more readily than homogeneous, ER-like topology~\cite{Tonjes2021CoherenceResonance,grinstein2005synchronous,mcgraw2003topology}.

Global degree structure, however, does not specify how connections are arranged locally. Cortical networks contain recurring higher-order subgraphs, or network motifs, including bidirectionally coupled pairs, feed-forward loops, and recurrent feed-forward loops~\cite{milo2002network,reigl2004search,song2005highly}. Such motifs provide distinct local pathways through which activity can be reinforced, filtered, or propagated~\cite{sporns2004motifs,hadjiabadi2023single}. Their dynamical effect therefore need not be determined by connection number or mean coupling strength alone, but may also depend on the particular arrangement of those connections.

Most studies have treated these structural scales separately. Motif analyses often consider small circuits in isolation~\cite{bonsel2022,jagdev2024,jagdev2024bi}, whereas studies of network coherence vary global topology without controlling local motif arrangement~\cite{masoliver2017coherence,baspinar2021coherence}. Once a motif is embedded, however, it receives input from the surrounding network and contributes to population activity. Its dynamics may therefore depend jointly on its internal wiring and on the topology in which it is embedded. It remains unclear whether motif-specific effects persist at the population level and how their contribution changes between homogeneous and hub-centred backgrounds.

We address this question using stochastic networks of Izhikevich (IZ) neurons~\cite{IZ} and compare four architectures: ER, motif-embedded ER (ERM), SF, and motif-embedded SF (SFM). ERM and SFM contain the same local motif skeleton but differ in their background inter-motif connectivity, while ER and SF provide matched controls without the imposed motifs. This design separates, as far as possible within these models, the contribution of local connection arrangement from that of global degree heterogeneity. We examine deterministic, noise-driven, and stimulus-driven dynamics; quantify coherence using the signal-to-noise ratio (SNR) of population- and motif-level firing rates; assess stimulus transmission using peristimulus time histograms (PSTHs) and cross-correlation with the input; and probe the roles of motif arrangement and hubs through rewiring and ablation.

All four architectures exhibit CR, but its magnitude and optimal noise intensity depend on network structure. Motif embedding increases noise-induced coherence and lowers the optimal noise intensity in both ER- and SF-based networks, with a larger proportional gain in the ER background. SF-based networks nevertheless attain substantially greater absolute coherence and produce earlier, larger, and more temporally concentrated responses to external stimulation. The corresponding transmission differences between motif-embedded networks and their non-motif controls are modest relative to trial-to-trial variability. Rewiring shows that the effect of local structure depends on the specific arrangement of motif connections rather than on strong local coupling alone, while targeted ablation shows that hub integrity contributes to the SF advantage. Together, these results indicate that local motif arrangement and global degree structure make distinct, interacting contributions to coherent population dynamics.

The remainder of this paper is organized as follows. Sec.~\ref{sec:model_and_methods} introduces the four network structures, the mathematical model, and methods. Sec.~\ref{sec:deterministic_dynamics} characterizes deterministic excitability; Secs.~\ref{sec:noise_induced_dynamics}--\ref{sec:noise_induced_coherence} establish noise-induced coherence resonance; Secs.~\ref{sec:inter_motif_connectivity}--\ref{subsec:dual_effects} trace its dependence on inter- and intra-motif connectivity; Sec.~\ref{sec:motif_rewiring} isolates the influence of specific motifs through rewiring; and Secs.~\ref{sec:information_transmission}--\ref{sec:ablation} assess signal transmission and the role of hubs through targeted ablation. Sec.~\ref{sec:discussion} discusses their implications.

\section{Model and Methods}
\label{sec:model_and_methods}

\subsection{Network structure}
\label{subsec:network_structure}

\begin{figure}[t]
    \centering
    \includegraphics[width=1\linewidth]{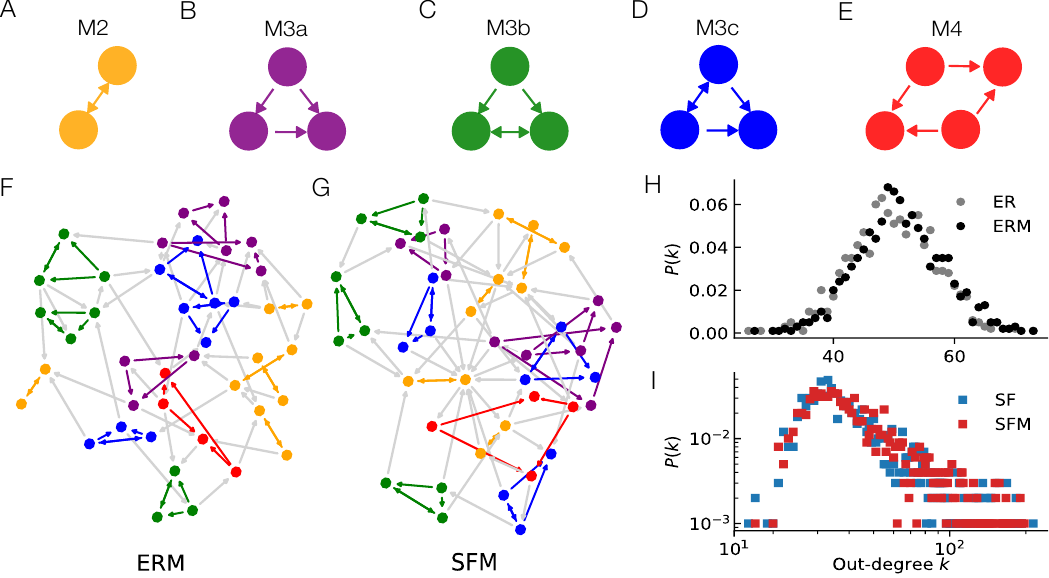}
    \caption{Motif structure and degree distribution. (A--E) motifs: bidirectionally coupled pair (M2, yellow), feed-forward loop (FFL; M3a, purple), type-1 recurrent FFL (RFFL; M3b, green), type-2 RFFL (M3c, blue), bi-parallel motif (M4, red). (F,G) Excitatory 41-neuron subgraphs of motif-embedded Erd\H{o}s-R\'enyi (ERM) and motif-embedded scale-free (SFM) networks; intra-motif edges coloured, inter-motif edges grey. (H,I) Out-degree distributions for ER/ERM (grey/black) and SF/SFM (blue/red) networks.}
    \label{fig:1}
\end{figure}

We consider four networks on $N=1000$ neurons: ER, ERM, SF, and SFM. In the ERM and SFM networks, motifs are realized as directed excitatory connections between neurons within a motif instance (intra-motif connections), embedded in a background of weaker random connections linking distinct motifs (inter-motif connections). Excitatory subgraphs of the ERM and SFM networks are shown in Fig.~\ref{fig:1}F,G, where coloured and grey edges denote intra- and inter-motif connections, respectively. The motifs considered are over-prevalent in cortical networks~\cite{milo2002network,reigl2004search,song2005highly}: M2, a bidirectionally coupled pair (Fig.~\ref{fig:1}A); M3a, a three-neuron feed-forward loop (FFL; Fig.~\ref{fig:1}B); M3b and M3c, two distinct three-neuron recurrent FFLs (RFFLs; Fig.~\ref{fig:1}C,D); and M4, a four-neuron bi-parallel (Fig.~\ref{fig:1}E).

Precluding autapses, inter-motif connections may form between any neuron pair. In the ER/ERM network, these edges are sampled independently with probability $p$. We consider two connection probabilities: $p_\mathrm{inh}$ and $p_\mathrm{exc}$, the probabilities that an inhibitory or excitatory neuron connects to any neuron. For simplicity we set $p=p_\mathrm{exc}=p_\mathrm{inh}$ unless otherwise stated. The out-degree therefore satisfies $$ \mathbb{P}[k_i=k] \approx \binom{N-1}{k}p^k(1-p)^{N-1-k}; \quad \mathbb{E}[k_i] \approx p(N-1).$$ The SF/SFM inter-motif edges are generated by a directed Barab\'asi-Albert preferential attachment rule with parameter $m$, yielding a heavy-tailed out-degree distribution $$ \mathbb{P}[k_i=k]\sim k^{-\gamma}; \quad \mathbb{E}[k_i]\approx m; \quad \gamma\approx 3.$$

The ERM and SFM networks share an identical motif skeleton (intra-motif edges), but their background connectivity is random. For parity, we fix $m=\lceil p(N-1)\rceil$, such that the SFM and ERM networks have an equal number of synapses.

For a fair comparison between the motif-embedded networks and their controls---SF vs.\ SFM, and ER vs.\ ERM---we adjust the shape parameters ($\overline{p}\geq p$, $\overline{m}\geq m$) to compensate for the absence of intra-motif edges, so all networks contain the same number of synapses. Moreover, the network degree distributions are primarily determined by inter-motif edges; intra-motif edges alter them negligibly. Any ERM or SFM neuron has at most two outgoing intra-motif edges but an expected $\lceil p(N-1)\rceil = m$ inter-motif edges, which for $N,\,p = 1000,\,0.05$ is $\approx 50$. Consequently, the empirical out-degree distributions of the ERM and SFM networks overlap with those of the ER and SF networks (Figs.~\ref{fig:1}H,I).

\subsection{Mathematical model}
\label{subsec:mathematical_model}
Neuron dynamics follow an adapted IZ model~\cite{IZ}. Each neuron evolves according to
\begin{align}
\mathrm{d}v_i(t) &= \left(0.04v_i^2 + 5v_i + 140 - u_i + I_i + I_{\mathrm{syn}}^{(i)}\right)\,\mathrm{d}t + D\,\mathrm{d}W_i(t), \label{v} \\
\mathrm{d}u_i(t) &= a\left(bv_i-u_i\right)\,\mathrm{d}t, \label{u}
\end{align}
where $v_i$ is the membrane potential and $u_i$ the recovery variable, subject to the auxiliary threshold-and-reset condition
\begin{equation}
\label{reset}
\text{if } v_i(t)\geq v_{\mathrm{th}}, \text{ then } v_i\leftarrow c,\quad u_i\leftarrow u_i+d.
\end{equation}
$I_i$ is a heterogeneous applied current, $I_i=\max\{0,Z_i\}$ with $Z_i\sim\mathcal{N}(\mu_I,\sigma_I^2)$; $\mu_I$ and $\sigma_I$ are control parameters. $D\,\mathrm{d}W_i(t)$ is an intrinsic noise term, where $D$ sets the noise level and $W_i(t)$ is a standard Wiener process, independent across neurons.

The synaptic current incoming to neuron $i$,
$I_{\mathrm{syn}}^{(i)}(t)= -I_{\mathrm{inh}}^{(i)}(t)+I_{\mathrm{intra}}^{(i)}(t)+I_{\mathrm{inter}}^{(i)}(t)$, consists of inhibitory and excitatory (intra- and inter-motif) streams. Each spike delivers a delta pulse to an exponentially decaying synaptic current:
\begin{align}
\frac{\mathrm{d}I_{\mathrm{inh}}^{(i)}}{\mathrm{d}t}
&= -\frac{I_{\mathrm{inh}}^{(i)}}{\tau_{\mathrm{inh}}}
+w_{\mathrm{inh}}\sum_{j \in \mathcal{P}^{(i)}_{\mathrm{inh}}}
\sum_{\tau^{(f)}_j \in \mathcal{F}^{(j)}}
\delta\!\left(t-\tau^{(f)}_j -\tau^{(d)}_{\mathrm{inh}}\right);
\label{eq:Iinh}
\\
\frac{\mathrm{d}I_{\mathrm{intra}}^{(i)}}{\mathrm{d}t}
&= -\frac{I_{\mathrm{intra}}^{(i)}}{\tau_{\mathrm{intra}}}
+w_{\mathrm{intra}}\sum_{j \in \mathcal{P}^{(i)}_{\mathrm{intra}}}
\sum_{\tau^{(f)}_j \in \mathcal{F}^{(j)}}
\delta\!\left(t-\tau^{(f)}_j - \tau^{(d)}_{\mathrm{intra}}\right);
\label{eq:Iintra}
\\
\frac{\mathrm{d}I_{\mathrm{inter}}^{(i)}}{\mathrm{d}t}
&= -\frac{I_{\mathrm{inter}}^{(i)}}{\tau_{\mathrm{inter}}}
+w_{\mathrm{inter}}\sum_{j \in \mathcal{P}^{(i)}_{\mathrm{inter}}}
\sum_{\tau^{(f)}_j \in \mathcal{F}^{(j)}}
\delta\!\left(t-\tau^{(f)}_j - \tau^{(d)}_{\mathrm{inter}}\right).
\label{eq:Iinter}
\end{align}
Here $\tau_{\mathrm{inh}}$, $\tau_{\mathrm{intra}}$, $\tau_{\mathrm{inter}}$ are synaptic decay time constants; $w_{\mathrm{inh}}$, $w_{\mathrm{intra}}$, $w_{\mathrm{inter}}$ are synaptic weights; and $\mathcal{P}^{(i)}_{\mathrm{inh}}$, $\mathcal{P}^{(i)}_{\mathrm{intra}}$, $\mathcal{P}^{(i)}_{\mathrm{inter}}$ are the sets of inhibitory, intra-motif, and inter-motif presynaptic neurons for neuron $i$. For each presynaptic neuron $j$, $\mathcal{F}^{(j)}$ is its set of spike times and $\tau_j^{(f)}$ the $f$th such time. The delays $\tau^{(d)}_{\mathrm{inh}}$, $\tau^{(d)}_{\mathrm{intra}}$, $\tau^{(d)}_{\mathrm{inter}}$ are treated as random variables (Table~\ref{tab:params}).

As $I_{\mathrm{intra}}^{(i)}=0$ in non-motif networks, we define an effective inter-motif coupling strength as a weighted average:
\begin{equation}
\label{w_bar}
\overline w_{\mathrm{inter}}
=
\frac{
w_{\mathrm{inter}}\,\mathbb{E}[|\mathcal{P}^{(i)}_{\mathrm{inter}}|]
+
w_{\mathrm{intra}}\,\mathbb{E}[|\mathcal{P}^{(i)}_{\mathrm{intra}}|]
}{
\mathbb{E}[|\mathcal{P}^{(i)}_{\mathrm{inter}}|]
+
\mathbb{E}[|\mathcal{P}^{(i)}_{\mathrm{intra}}|]
},
\end{equation}
where $i$ indexes an excitatory neuron in a motif-embedded network. Setting $w_{\mathrm{intra}}=1$ and $w_{\mathrm{inter}}=0.25$, for example, yields $\overline w_{\mathrm{inter}}\approx 0.266$ for the corresponding non-motif control.

Excitatory synapses carry spike-timing-dependent plasticity (STDP). Each excitatory synapse holds pre- and postsynaptic traces, $A_{\mathrm{pre}}$ and $A_{\mathrm{post}}$, decaying with time constants $\tau_{\mathrm{pre}}$ and $\tau_{\mathrm{post}}$. Each pre- or postsynaptic spike increments the corresponding trace ($A_{\mathrm{pre}} \to A_{\mathrm{pre}} + \mathrm{d}A_{\mathrm{pre}}$, $A_{\mathrm{post}} \to A_{\mathrm{post}} + \mathrm{d}A_{\mathrm{post}}$). At a postsynaptic spike the weight is potentiated by $A_{\mathrm{pre}}$; at a presynaptic spike it is depressed by $A_{\mathrm{post}}$, with $w_{ij} \leftarrow \mathrm{clip}(w_{ij}+A_{\mathrm{pre/post}},0,w_{\max})$. The weight $w_{ij}$ is initialized to either $w_{\mathrm{intra}}$, $w_{\mathrm{inter}}$, or $\overline w_{\mathrm{inter}}$.

Parameters $a,b,c,d$ govern the IZ neuron's firing pattern. Inhibitory neurons are fast-spiking ($a=0.1$, $b=0.2$, $c=-65$, $d=2$)~\cite{IZ}. Excitatory neurons are heterogeneous, with $a=0.02$, $b=0.2$, $c_i=-65+15r_i^2$, $d_i=8-6r_i^2$, $r_i\sim\mathcal{U}[0,1]$, where $r_i=0$ and $r_i=1$ correspond to regular-spiking and chattering regimes, respectively. The quadratic weighting biases the population toward regular-spiking cells---the dominant excitatory type in the cortex~\cite{IZ, connors1990intrinsic}.

\subsection{Methods}
\label{subsec:methods}
Simulations are implemented in Python using Brian2~\cite{stimberg2019}. Unless otherwise stated, all networks comprise $N=1000$ neurons and results are averaged over at least 30 independent trials, with randomized initial conditions and inter--motif connectivity; Sec.~\ref{sec:motif_rewiring} instead uses a smaller network ($N=500$) to make the effect of rewiring more apparent. Eqs.~\eqref{v}--\eqref{eq:Iinter} are integrated by the Euler--Maruyama method with time step $\mathrm{d}t=0.5$~ms over a total duration $T=2000$~ms, discarding an initial transient of length $t_0=200$~ms. Initial conditions for each neuron and trial are set as $v_i(0)=c_i Q$, $Q\sim\mathcal{U}[0,1]$, and $u_i(0)=bv_i(0)$, following~\cite{IZ}. Simulation output is exported and analyzed in MATLAB. Statistical significance, where reported, is assessed by a permutation test~\cite{holt2023permutation,liu2024permutation} with $10^4$ random permutations. Additional parameters are listed in Table~\ref{tab:params}.

\section{Results}
\label{sec:results}
\subsection{Deterministic dynamics}
\label{sec:deterministic_dynamics}
We first characterize network dynamics in the deterministic regime ($D=0$), in which neurons receive only the input current ($I_i$) and synaptic current ($I_\mathrm{syn}^{(i)}$). The network response is summarized by the per-neuron population firing rate,
\begin{equation}
\label{Eq:f}
f=
\frac{1}{N(T-t_0)}
\sum_{i=1}^N
\sum_{t_0 \leq t \leq T}
\mathbbm{1} [v_i(t)\geq v_{\mathrm{th}}],
\end{equation} 
where $\mathbbm{1}$ is the indicator function.

We consider ERM and SFM networks with $w_{\mathrm{inh}}=2$, $w_{\mathrm{inter}}=0.25$, $w_{\mathrm{intra}}=1$, and $p_{\mathrm{inh}}=p_{\mathrm{exc}}=0.05$ (ERM), $m=50$ (SFM). The corresponding controls use matched effective parameters: $\overline w_{\mathrm{inter}}\approx 0.266$ and $\overline p_{\mathrm{exc}}\approx0.0511$ for ER, and $\overline m=51$ for SF.

\begin{figure}[t]
    \centering
    \includegraphics[width=1\linewidth]{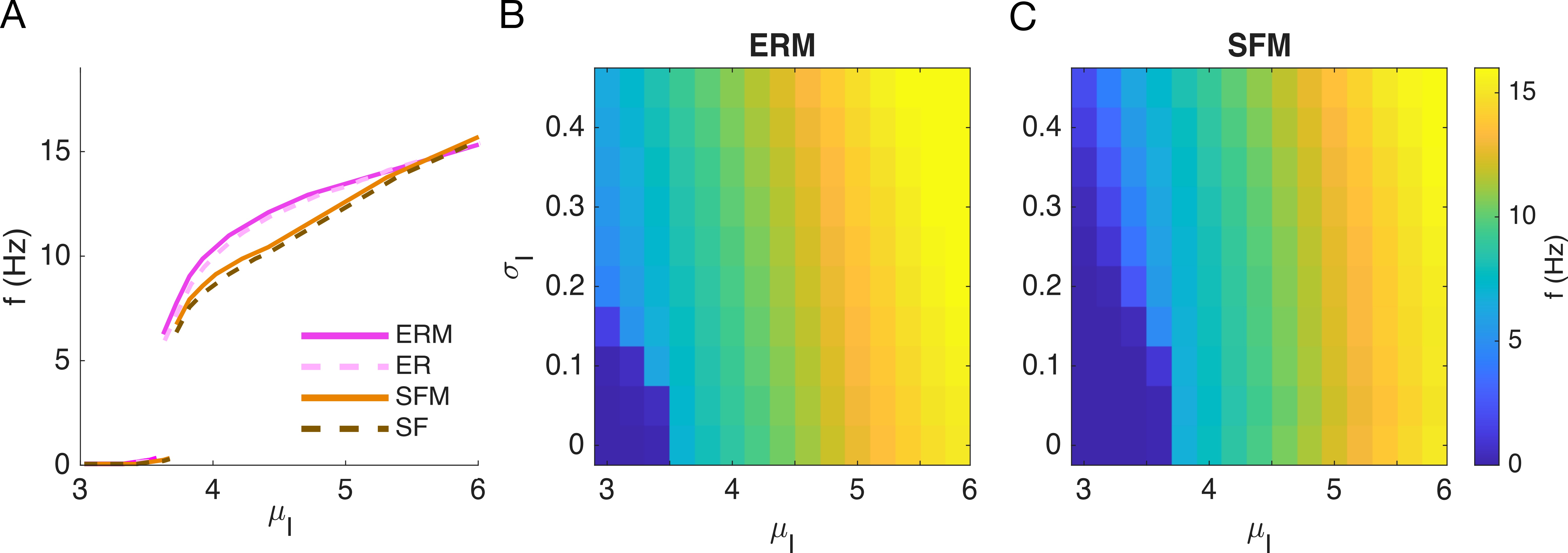}
    \caption{Deterministic firing-rate responses. (A) $f$-$I$ curves vs.\ mean input current $\mu_I$: ERM (solid pink), ER (dotted pink), SFM (solid orange), SF (dotted brown). (B,C) $f$-$I$ surfaces over $(\mu_I,\sigma_I)$ for ERM (B) and SFM (C).}
    \label{fig:2}
\end{figure}

The $f$-$I$ curves obtained by varying $\mu_I$ with $\sigma_I=0$ are shown in Fig.~\ref{fig:2}A. For weak inputs ($\mu_I \lesssim 3.55$), all networks remain quiescent. As $\mu_I$ increases, the networks undergo an abrupt transition to sustained activity, which we interpret as an excitation threshold. For the ER and ERM networks this transition occurs near $\mu_I \approx 3.6$, where the firing rate jumps from $\approx 0$ to $\approx 6.6$~Hz; for the SF and SFM networks it occurs near $\mu_I \approx 3.7$, where $f$ jumps from $\approx 0$ to $\approx 6.7$~Hz. Thus, although the SF and SFM networks have a larger excitation threshold, the firing rate immediately above threshold is similar across all networks
($\approx 6.6$-$6.7$~Hz).

When $3.7 \leq \mu_I \leq 5.35$, however, the ER and ERM networks reach higher $f$ than the SF and SFM networks: at $\mu_I \approx 4.2$, for example, the ER and ERM firing rates are $\approx 10.5$ and $\approx 10.7$~Hz, respectively, whereas the SF and SFM rates are $\approx 9.3$ and $\approx 9.9$~Hz. These differences diminish as $\mu_I$ increases beyond $\approx 4.3$, until the curves coincide near $\mu_I \approx 5.35$, where $f \approx 14.4$~Hz for all networks. 

Varying both $\mu_I$ and $\sigma_I$ in the motif-embedded networks, we examine $f$ over the $(\mu_I,\sigma_I)$ parameter space (Fig.~\ref{fig:2}B,C). Increasing $\sigma_I$ lets some neurons spike even while $\mu_I$ remains below the deterministic excitation threshold of Fig.~\ref{fig:2}A. Increasing $\sigma_I$ at fixed $\mu_I$ generally increases $f$, though the effect is weaker for large $\mu_I$ than for $\mu_I<3.6$, where the excitation threshold falls as $\sigma_I$ increases. The ERM network is more readily excited than the SFM network: for $\sigma_I>0.2$, the ERM network is never quiescent, unlike the SFM network. This difference appears in the region of the heat maps below the deterministic
excitation threshold.

\subsection{Noise-induced dynamics}
\label{sec:noise_induced_dynamics}
\begin{figure}[t]
    \centering
    \includegraphics[width=1\linewidth]{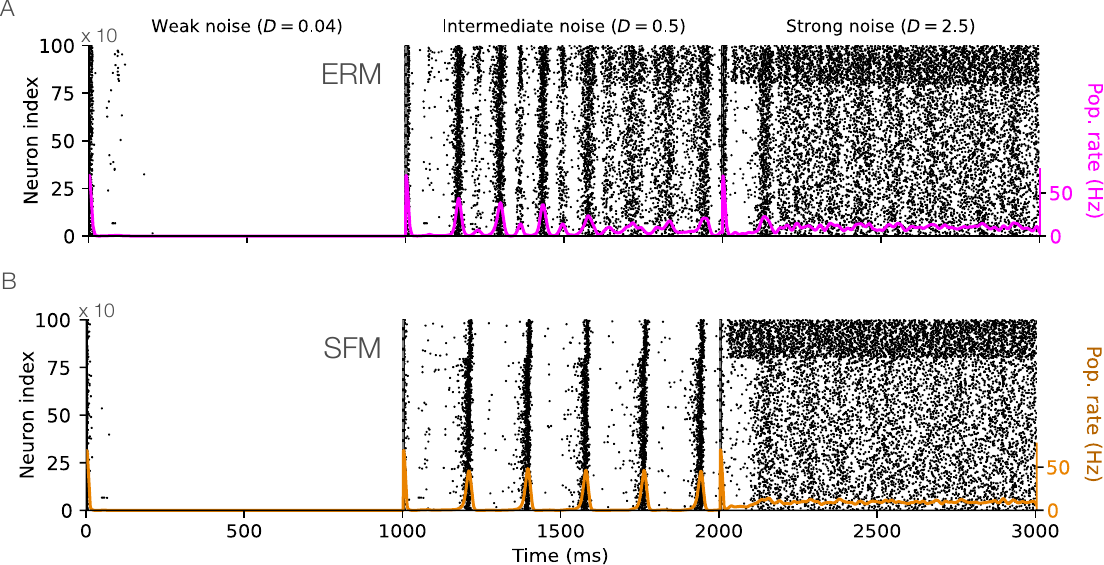}
    \caption{Noise-induced population dynamics in the quiescent regime. Raster plots and population-averaged firing rate (ERM: pink; SFM: orange) vs.\ noise intensity $D$ for (A) ERM and (B) SFM networks.}
    \label{fig:3}
\end{figure}

In the quiescent regime ($\mu_I=3.4$, $\sigma_I=0.1$), where $I_i$ alone drives only transient activity, the addition of intrinsic noise can induce tonic spiking. Fig.~\ref{fig:3} shows raster plots and PSTHs for the ERM (Fig.~\ref{fig:3}A) and SFM (Fig.~\ref{fig:3}B) networks across weak, intermediate, and strong noise regimes.

In the weak-noise regime (e.g., $D=0.04$), activity in both networks is transient, and population firing rates remain near zero, reflecting that $D$ is insufficient to drive tonic spiking. The ERM network shows a slightly higher frequency of transient spikes than the SFM network, consistent with Fig.~\ref{fig:2}.

At intermediate noise intensities (e.g., $D=0.5$), both networks transition to tonic spiking. The SFM network produces sharper, higher PSTH peaks, the ERM network broader, lower ones; the raster plots show the same contrast, with SFM firing highly synchronized and ERM dynamics more dispersed. The inter-spike intervals (ISIs) of the SFM network are correspondingly larger and more regular than those of the ERM network, consistent with the lower firing rate of the SF and SFM networks (Fig.~\ref{fig:2}).

In the strong-noise regime (e.g., $D=2.5$), this temporal structure breaks down in both networks: the raster plots desynchronize, with ill-defined PSTH peaks and irregular ISIs. Coherence therefore rises then falls with $D$, peaking at intermediate noise---the signature of CR.

\subsection{Noise-induced coherence}
\label{sec:noise_induced_coherence}

To quantify these differences, we use the (trial-averaged) signal-to-noise ratio
(SNR), measured about the dominant spectral peak,
\begin{equation}
\label{SNR}
\mathrm{SNR}(D)
=
\sqrt{\frac{\frac{1}{M}\sum_{m \in \mathcal{W}} \left|\mathrm{DFT}\{r\}[m]\right|^2}{D^2}},
\end{equation}
where $\mathrm{DFT}\{r\}$ is the discrete Fourier transform of the population firing rate $r(t)=N^{-1}\sum_{i=1}^N x_i(t)$, with $x_i(t)=\sum_f \delta(t-\tau_i^{(f)})$ the spike train of neuron $i$; $r(t)$ is binned to form the PSTH. Here $\mathcal{W}$ is a window of $M$ frequency bins centred on the dominant spectral peak $m^\ast = \arg\max_m |\mathrm{DFT}\{r\}[m]|$. Larger SNR values indicate that the power spectrum is more strongly concentrated about this peak, and hence more temporally structured---that is, the response has greater spike-timing coherence.

\begin{figure}[t]
    \centering
    \includegraphics[width=1\linewidth]{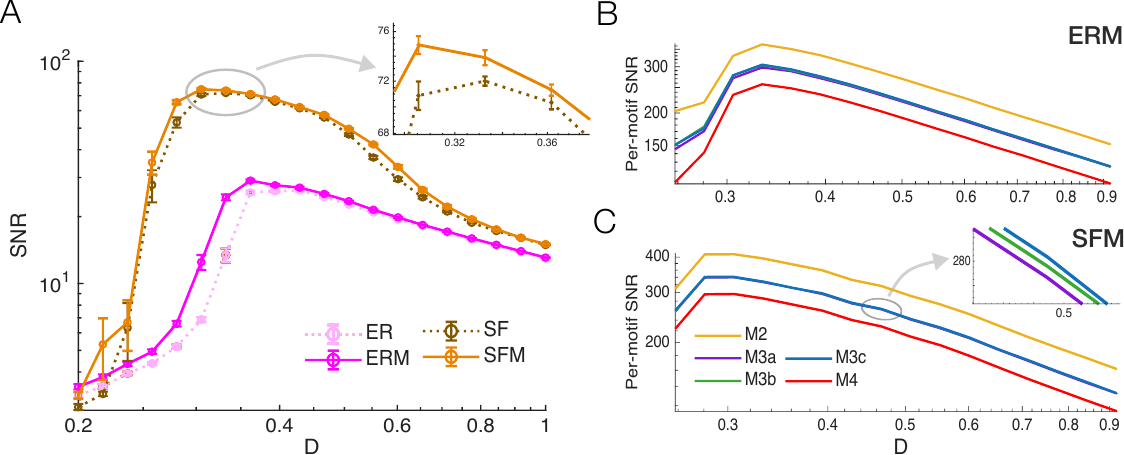}
    \caption{Network- and motif-level coherence. (A) SNR vs.\ $D$: ER (dotted light pink), ERM (solid pink), SF (dotted brown), SFM (solid orange); inset details the SF/SFM comparison near their peaks (error bars: trial-to-trial s.d.). (B,C) Motif-level SNR vs.\ $D$ for ERM (B) and SFM (C); inset in (C) resolves the near-coincident M3a/M3b/M3c curves.}
    \label{fig:4}
\end{figure}

We evaluate the SNR across all networks as a function of $D$ (Fig.~\ref{fig:4}A), where dotted light pink, solid pink, dotted brown, and solid orange curves denote the ER, ERM, SF, and SFM networks, respectively. For each network, the SNR depends non-monotonically on $D$, attaining a peak value $\mathrm{SNR}^*=\max_D\{\mathrm{SNR}(D)\}$ at an optimal (intermediate) noise level $D^*=\arg\max_D\{\mathrm{SNR}(D)\}$; all networks therefore exhibit CR.

In both ER- and SF-based topologies, motif embedding shifts the SNR curve upward: $\mathrm{SNR}^*\approx 26.2$ and $29$ for the ER and ERM networks, and $\mathrm{SNR}^*\approx 72.0$ and $74.1$ for the SF and SFM networks. The proportional gain is larger for the ER-based topology ($\approx 11\%$, against $\approx 3\%$ for SF). Motif embedding also lowers the optimal noise intensity in both cases, from $D^*\approx 0.428$ to $0.36$ for ER/ERM and from $D^*\approx 0.33$ to $0.30$ for SF/SFM. This leftward shift in $D^*$ is larger for ER/ERM, concentrating the ER motif benefit at weak noise ($D<0.4$).

To assess whether these differences are statistically significant, we apply a permutation test~\cite{holt2023permutation,liu2024permutation}. We compare the noise-averaged SNR, $\langle\mathrm{SNR}\rangle_D=|\mathcal{D}|^{-1}\sum_{D\in\mathcal{D}}\mathrm{SNR}(D)$, between each motif-embedded network and its non-motif control, where $\mathcal{D}$ is the set of noise intensities considered. Both inequalities $\langle\mathrm{SNR}\rangle_D^{\mathrm{ERM}}>\langle\mathrm{SNR}\rangle_D^{\mathrm{ER}}$ and $\langle\mathrm{SNR}\rangle_D^{\mathrm{SFM}}>\langle\mathrm{SNR}\rangle_D^{\mathrm{SF}}$ are supported ($p$-value $\approx 0.0005$). The inset of Fig.~\ref{fig:4}A shows this comparison directly, with error bars giving the trial-to-trial standard deviation of the SF and SFM curves.

As the ERM and SFM networks contain distinct motif types, we further consider the motif-level SNR, defined as the SNR computed from the PSTH of each motif instance, averaged across all instances of that type. Results are presented in Figs.~\ref{fig:4}B,C for the ERM and SFM networks, respectively. In both networks the M2 motif (yellow) attains the largest SNR across all $D$, and M4 (red) the lowest; the M3a, M3b, and M3c curves appear coincident at the scale of the main panels, though the inset of Fig.~\ref{fig:4}C reveals
$\langle\mathrm{SNR}\rangle_D^{\mathrm{M3c}} \gtrsim \langle\mathrm{SNR}\rangle_D^{\mathrm{M3b}} > \langle\mathrm{SNR}\rangle_D^{\mathrm{M3a}}$. The M3c-M3b difference is not statistically significant ($p$-value $\approx 0.15$ for ERM, $\approx 0.13$ for SFM), whereas the M3b-M3a difference is supported ($p$-value $\leq 0.019$ in both networks).

As the ERM and SFM networks share an identical motif skeleton, differences in motif-level SNR between them reflect the background connectivity alone. At weak noise ($D<0.3$), the per-motif SNR is higher in SFM than in ERM for every motif type: the same local circuits fire more coherently on an SF topology than on an ER one. This gap narrows as $D$ increases past $D^*$. SF topology thus amplifies local-motif coherence most strongly at weak noise, where it supplies coordinated background drive; at strong noise, the noise dominates and background connectivity matters less.

Moreover, the network- and motif-level views need not agree. At intermediate noise ($D\approx 0.4$), the SFM and ERM networks differ substantially in network-level SNR, yet their per-motif SNRs are nearly equal. Conversely, at strong noise ($D\approx 0.9$), the network-level curves converge while the separation between motif types---e.g., $\mathrm{SNR}_\mathrm{M2}>\mathrm{SNR}_\mathrm{M4}$---persists. The motifs thus behave as coherent local clusters in their own right. We return to this point in Sec.~\ref{sec:discussion}.

\subsection{Effects of inter-motif connectivity}
\label{sec:inter_motif_connectivity}

We next consider how coherence depends on the density and strength of the background connectivity. 

First we vary the excitatory inter-motif connection probability $p_{\mathrm{exc}}$, with the SF/SFM matched as $m=\lceil p_{\mathrm{exc}}(N-1)\rceil$. In the ER and SF controls, the effective parameters $\overline{p}_{\mathrm{exc}}$ and $\overline{m}$ are adjusted to preserve the total synapse count, while $p_\mathrm{inh}=0.05$ is fixed.

\begin{figure}[t]
\centering
    \includegraphics[width=1\linewidth]{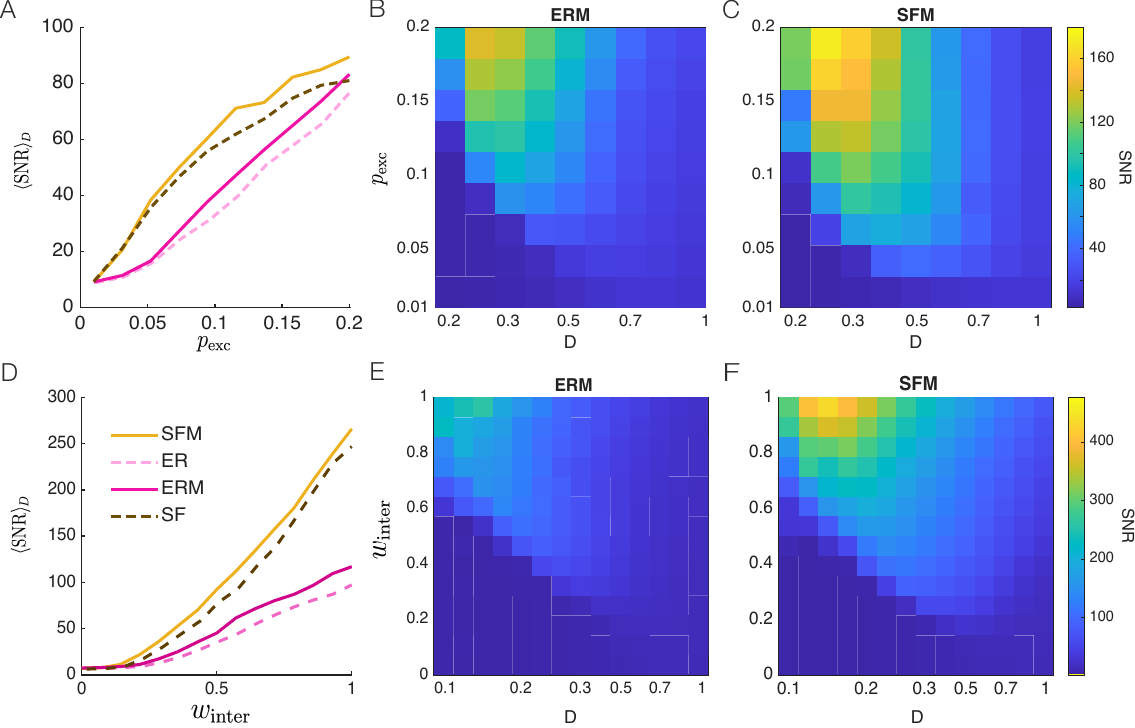}
    \caption{Effects of inter-motif connectivity and coupling strength on coherence. (A) $\langle\mathrm{SNR}\rangle_D$ vs.\ connection probability $p_{\mathrm{exc}}$: ER, ERM, SF, SFM. (B,C) SNR vs.\ $(D,p_{\mathrm{exc}})$ for ERM (B) and SFM (C). (D) $\langle\mathrm{SNR}\rangle_D$ vs.\ inter-motif coupling strength $w_{\mathrm{inter}}$: ER, ERM, SF, SFM. (E,F) SNR vs.\ $(D,w_{\mathrm{inter}})$ for ERM (E) and SFM (F).}
    \label{fig:5}
\end{figure}

Fig.~\ref{fig:5}A shows $\langle\mathrm{SNR}\rangle_D$ as a function of $p_{\mathrm{exc}}$ (which sets $m$) for all networks. $\langle\mathrm{SNR}\rangle_D$ increases monotonically with $p_{\mathrm{exc}}$ across all networks. For $p_{\mathrm{exc}}\leq 0.05$ the rise is gradual in the ER and ERM networks but much steeper in SF and SFM. In this region the motif-embedded networks are nearly coincident with their controls, and the SF and SFM curves lie above those of the ER and ERM networks.

For $p_{\mathrm{exc}}>0.05$ the SF and SFM curves become concave, while the ER and ERM curves remain approximately linear. As $p_{\mathrm{exc}}$ increases further, the motif-embedded networks diverge from their controls, the difference peaking at $p_{\mathrm{exc}}=0.2$. At the same time the ER/ERM and SF/SFM curves converge: as connectivity approaches saturation, the edge arrangement---random (ER) or preferential (SF)---matters less for population dynamics, though the motif skeleton still confers some coherence benefit. (For $p_{\mathrm{exc}}>0.2$ the networks are
no longer excitable, transitioning to tonic spiking.)

Figs.~\ref{fig:5}B,C display the SNR as a function of $D$ and $p_{\mathrm{exc}}$ for the ERM and SFM networks, respectively. The two heat maps share the same qualitative structure: for each fixed $p_{\mathrm{exc}}$, the SNR varies non-monotonically with $D$, attaining a maximum at $D^*$, and the peak broadens over a wider noise range as $p_{\mathrm{exc}}$ increases. Quantitatively, however, the SFM network attains larger SNR over a broad region of the parameter space, most notably for $0.05 \lesssim p_{\mathrm{exc}} \lesssim 0.2$ and $0.2 \lesssim D \lesssim 0.7$. In both panels the low-SNR (dark) region recedes as $p_{\mathrm{exc}}$ increases: higher connectivity lowers the noise intensity needed to drive sustained activity.

We next vary the inter-motif coupling strength $w_{\mathrm{inter}}$. Fig.~\ref{fig:5}D shows $\langle\mathrm{SNR}\rangle_D$ as a function of $w_{\mathrm{inter}}$ for the ERM and SFM networks, and of $\overline{w}_{\mathrm{inter}}$ for the ER and SF networks (for brevity the figure uses only the label $w_{\mathrm{inter}}$, though equivalent changes are made to both parameters). For small $w_{\mathrm{inter}}$ (e.g., $w_{\mathrm{inter}}<0.13$), the $\langle\mathrm{SNR}\rangle_D$ curves coincide, likely because the networks cannot sustain spiking in this regime. As $w_{\mathrm{inter}}$ increases beyond this threshold, the two topologies diverge markedly: SF and SFM show an exponential-like rise, while ER and ERM increase almost linearly with a shallow slope.

Furthermore, the ordering of $\langle\mathrm{SNR}\rangle_D$ is preserved across $w_{\mathrm{inter}}$: even at large $w_{\mathrm{inter}}$, the inequality $\langle\mathrm{SNR}\rangle_D^\mathrm{SFM}>\langle\mathrm{SNR}\rangle_D^\mathrm{SF}>\langle\mathrm{SNR}\rangle_D^\mathrm{ERM}>\langle\mathrm{SNR}\rangle_D^\mathrm{ER}$ holds for all $w_{\mathrm{inter}}\gtrsim 0.2$ ($p$-value $\leq 0.002$ at $w_{\mathrm{inter}}=1$).

Fig.~\ref{fig:5}E,F shows SNR as a function of $D$ and $w_\mathrm{inter}$ for the ERM (E) and SFM (F) networks. The two heat maps differ quantitatively---the SFM SNR is almost everywhere greater than the ERM SNR---but are qualitatively similar over the $(D, w_\mathrm{inter})$ space, both displaying CR throughout. As $w_\mathrm{inter}$ increases, $\mathrm{SNR}^*$ increases while $D^*$ decreases.

Increasing $p_{\mathrm{exc}}$ and $w_{\mathrm{inter}}$ have opposite effects: $p_{\mathrm{exc}}$ draws the SF/SFM and ER/ERM curves together, whereas $w_{\mathrm{inter}}$ pushes them apart. Higher connection density benefits the ER/ERM networks more, converging the two topologies in coherence; higher coupling strength benefits the SF/SFM networks more, diverging them. This asymmetry is expected: increasing connection density is homogenizing at high $p_{\mathrm{exc}}$, as both topologies approach dense connectivity, whereas increasing coupling  strength amplifies the existing structural difference rather than erasing it. The benefit of motif embedding is largely unaffected by either background connection strength or density, except at weak values of either parameter, where network activity is sparse.

\subsection{Effects of intra-motif connection strength}
\label{subsec:intra_strength}
\begin{figure}[t]
    \centering
    \includegraphics[width=1\linewidth]{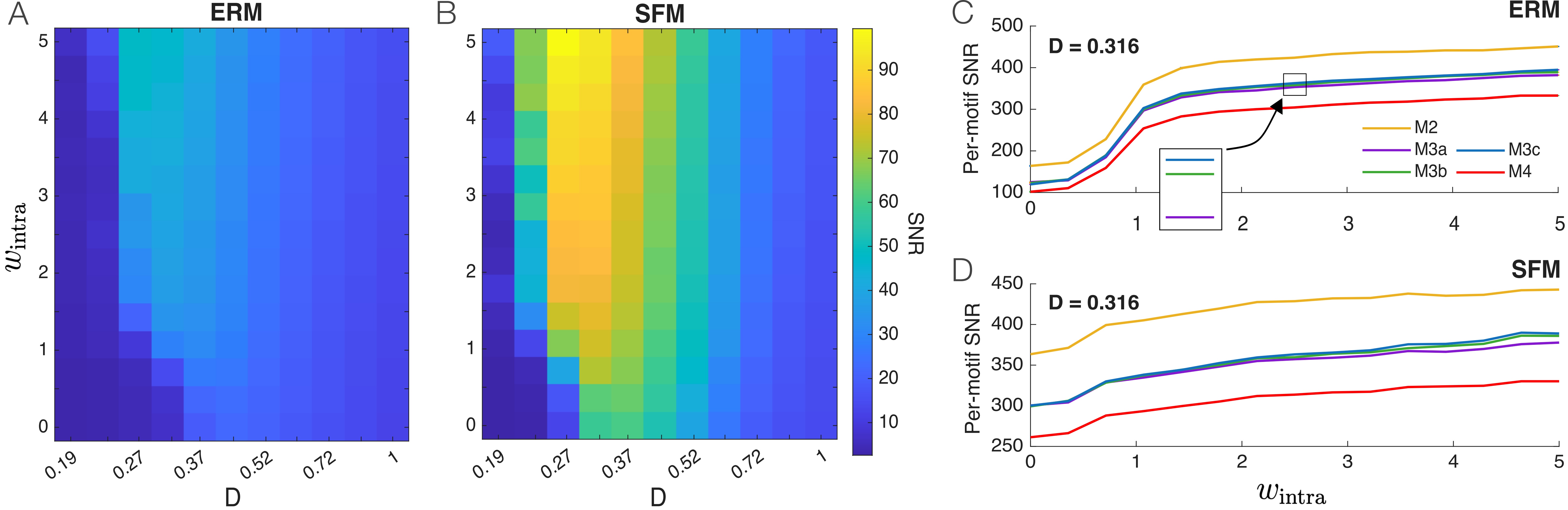}
    \caption{Effects of intra-motif coupling strength $w_\mathrm{intra}$ on SNR. (A,B) SNR vs.\ $(D,w_\mathrm{intra})$ for ERM (A) and SFM (B). (C,D) Per-motif SNR vs.\ $w_\mathrm{intra}$ at $D=0.316$ for ERM (C) and SFM (D); inset in (C) resolves the near-coincident M3a/M3b/M3c curves.}
    \label{fig:6}
\end{figure}

We quantify the effect of varying $w_\mathrm{intra}$ using the SNR (Eq.~\ref{SNR}). Fig.~\ref{fig:6}A,B shows the SNR over the $(D, w_\mathrm{intra})$ space for the ERM (A) and SFM (B) networks. The two heat maps are qualitatively similar: both display CR throughout the $w_\mathrm{intra}$ range, with a well-defined SNR maximum at intermediate $D$. The optimal noise intensity $D^*$ is shared between the two networks and shifts with $w_\mathrm{intra}$.

For sufficiently large $w_\mathrm{intra}$ (e.g., $w_\mathrm{intra}\gtrsim 1.5$), the optimal noise intensity is nearly constant at $D^*\approx 0.3$, whereas for smaller $w_\mathrm{intra}$ it shifts to larger values as $w_\mathrm{intra}$ decreases. In both networks the SNR increases with $w_\mathrm{intra}$ over the full range; quantitatively, however, the SFM network attains larger SNR than ERM. At $w_\mathrm{intra}=5$ the SFM $\mathrm{SNR}^*$ is roughly twice the ERM value (Fig.~\ref{fig:6}A,B).

We further examine the motif-specific effects of varying $w_{\mathrm{intra}}$. Figs.~\ref{fig:6}C,D show the per-motif SNR in the ERM (C) and SFM (D) networks as a function of $w_{\mathrm{intra}}$ at $D=0.316$. In both networks the ordering $\mathrm{SNR}_{\mathrm{M2}} > \mathrm{SNR}_{\mathrm{M3c}} > \mathrm{SNR}_{\mathrm{M3b}} > \mathrm{SNR}_{\mathrm{M3a}} > \mathrm{SNR}_{\mathrm{M4}}$ holds across the $w_\mathrm{intra}$ range (inset of Fig.~\ref{fig:6}C). At $w_{\mathrm{intra}} = 2.5$, for example, $\mathrm{SNR}_{\mathrm{M3c}}$ and $\mathrm{SNR}_{\mathrm{M3b}}$ both exceed $\mathrm{SNR}_{\mathrm{M3a}}$ in both networks ($p$-value $< 0.002$), whereas the difference $\mathrm{SNR}_{\mathrm{M3c}} > \mathrm{SNR}_{\mathrm{M3b}}$ is small and not statistically significant ($p$-value $\geq 0.17$ in both).

For small $w_{\mathrm{intra}}$ (e.g., $w_{\mathrm{intra}} < 1.167$; Fig.~\ref{fig:6}C,D), the per-motif SNR curves of the ERM and SFM networks differ appreciably: at $w_{\mathrm{intra}} = 0.5$, the M2 curve is $\approx 200$ for ERM and $\approx 375$ for SFM. As $w_{\mathrm{intra}}$ increases these differences diminish; both M2 curves plateau at $\approx 410$ for $w_{\mathrm{intra}} > 2$. All per-motif ERM curves rise sharply over $0.25 \leq w_{\mathrm{intra}} \leq 1$. This convergence is local only: in the network-level heat maps (Fig.~\ref{fig:6}A,B) the SFM SNR remains substantially larger than the ERM SNR over much of the parameter space. For sufficiently large $w_{\mathrm{intra}}$, then, individual motifs in both networks form highly coherent local clusters, even as network coherence stays strongly dependent on global structure.

This corroborates our findings in Fig.~\ref{fig:4}B,C. At weak intra-motif coupling the SF background supports motif coherence, giving the SFM network markedly higher per-motif SNR than ERM; at strong coupling the two networks' per-motif curves converge, the motifs sustain coherent spiking independently of the background---just as the per-motif curves converged at strong noise while the network-level curves remained apart.

\subsection{Dual effects of intra- and inter-motif connection strength}
\label{subsec:dual_effects}
Thus far we have considered the effects of $w_{\mathrm{intra}}$ and
$w_{\mathrm{inter}}$ independently; we now vary them jointly. As inter-motif synapses far outnumber intra-motif synapses, the weights $w_{\mathrm{intra}}$ and $w_{\mathrm{inter}}$ do not directly reflect the relative contribution of intra- and inter-motif coupling: the preceding networks have $\approx 50{,}000$ inter-motif synapses yet only $\approx 903$ intra-motif synapses. We therefore normalize each weight by the number of incoming synapses, replacing $w_{\mathrm{intra}}$ and $w_{\mathrm{inter}}$ in Eqs.~\eqref{eq:Iintra}-\eqref{eq:Iinter} with $w_{\mathrm{intra}}/|\mathcal{P}^{(i)}_{\mathrm{intra}}|$ and $w_{\mathrm{inter}}/|\mathcal{P}^{(i)}_{\mathrm{inter}}|$, respectively.

\begin{figure}[t]
    \centering
    \includegraphics[width=1\linewidth]{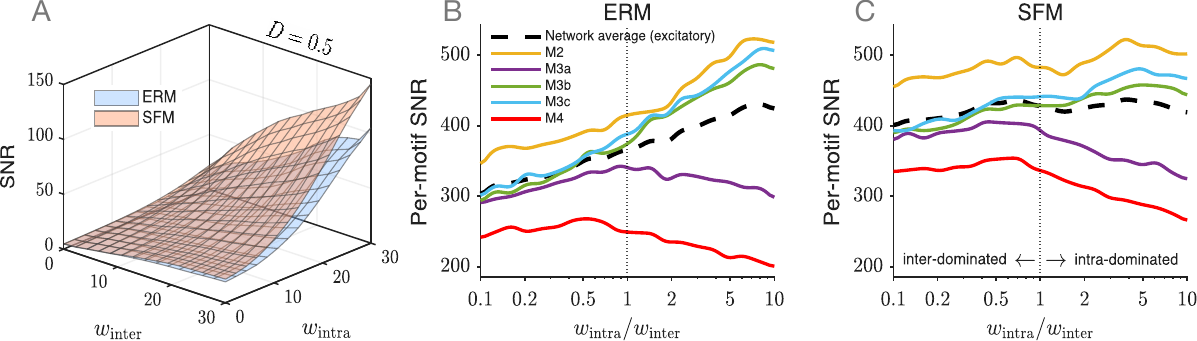}
    \caption{Dual effects of intra- and inter-motif coupling strength on SNR. (A)
    SNR vs.\ $(w_{\mathrm{intra}},w_{\mathrm{inter}})$ for $D=0.5$ for ERM (blue) and
    SFM (red). (B,C) Per-motif SNR vs.\ $w_{\mathrm{intra}}/w_{\mathrm{inter}}$ for
    ERM (B) and SFM (C) with fixed total weight $w_{\mathrm{intra}}+w_{\mathrm{inter}}= 60$.}
    \label{fig:7}
\end{figure}

For fixed $D=0.5$, Fig.~\ref{fig:7}A shows the SNR surface as a function of $w_{\mathrm{intra}}$ and $w_{\mathrm{inter}}$ for the ERM (blue) and SFM (red) networks. On both surfaces, the SNR increases with $w_{\mathrm{intra}}$ and $w_{\mathrm{inter}}$, reaching its maximum at the largest coupling strengths. The SFM surface lies above the ERM surface everywhere except in the weak-coupling regime ($w_\mathrm{inter}, w_\mathrm{intra} < 10$), where the surfaces coincide, likely due to sparse network activity.

Next we fix the total coupling strength, $w_{\mathrm{intra}}+w_{\mathrm{inter}}=60$,
and vary the ratio $w_{\mathrm{intra}}/w_{\mathrm{inter}}$ over $0.1\leq w_{\mathrm{intra}}/w_{\mathrm{inter}}\leq 10$ at $D=0.5$. Increasing this ratio redistributes coupling strength from inter- to intra-motif connections. Figs.~\ref{fig:7}B,C show the per-motif SNR, together with the excitatory-population SNR computed from the pooled PSTH of all excitatory neurons, for the ERM and SFM networks, respectively. In the inter-motif-dominated regime the SFM excitatory population (dashed black curve) has higher SNR than the ERM population, but the two converge as coupling shifts toward intra-motif connections. At $w_{\mathrm{intra}}/w_{\mathrm{inter}}=0.1$ the ERM and SFM populations have $\mathrm{SNR}\approx 300$ and $\approx 400$, respectively; at $w_{\mathrm{intra}}/w_{\mathrm{inter}}=10$ both approach $\mathrm{SNR}\approx 420$, with no significant difference ($p\approx 0.84$). The SFM advantage is thus largest under inter-motif-dominated coupling and diminishes as strength shifts toward intra-motif connections.

Motifs M2, M3b, and M3c respond similarly: their SNR increases as coupling shifts toward intra-motif connections, more steeply in ERM than in SFM. The M3b and M3c curves rise from $\mathrm{SNR}\approx 300$ in the inter-motif-dominated regime to $\approx 480$ and $\approx 500$, respectively, in the intra-motif-dominated regime; in SFM the corresponding curves rise from $\approx 390$ to $\approx 455$ and $\approx 465$. This greater sensitivity of the ERM motifs is consistent with the SF topology providing stronger support for coherent motif activity when intra-motif coupling is weak.

Motifs M3a and M4 show a qualitatively different dependence. Rather than increasing monotonically, their SNR reaches an interior maximum, then declines as coupling shifts further toward intra-motif connections. In SFM both peak near $w_{\mathrm{intra}}/w_{\mathrm{inter}}\approx 0.7$; in ERM the maxima occur near $\approx 1$ for M3a and $\approx 0.5$ for M4. M3a and M4 thus attain greatest coherence at balanced coupling ratios, in contrast to M2, M3b, and M3c, whose coherence increases as coupling becomes intra-motif-dominated.

\subsection{Motif rewiring}
\label{sec:motif_rewiring}

\begin{figure}[t]
\centering
\includegraphics[width=1\linewidth]{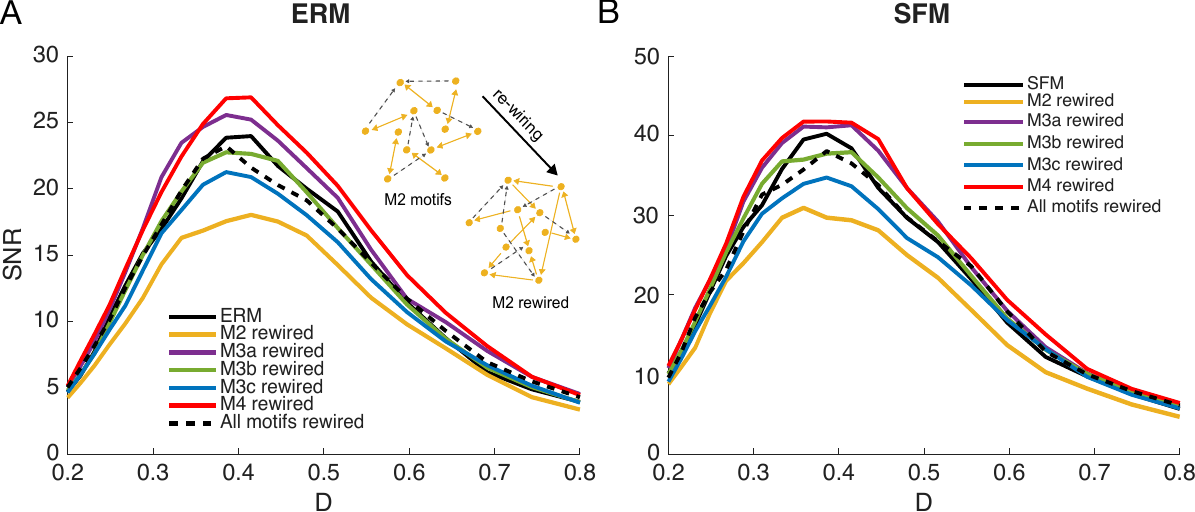}
\caption{Effects of motif rewiring on SNR ($p=0.02$, $w_\mathrm{intra}=1.75$, $w_\mathrm{inter}=0.5$; excitatory sub-network). (A,B) SNR vs.\ $D$ for rewired ERM (A) and SFM (B): intact (black), M2/M3a/M3b/M3c/M4 rewired (yellow/purple/green/blue/red), all motifs rewired (dashed black). Inset in (A): M2-motif rewiring schematic, showing intra-motif (yellow) and inter-motif (dashed grey) synapses before/after rewiring.}
\label{fig:8}
\end{figure}

In the preceding sections, we showed that motif-embedded networks are more coherent than their non-motif controls. This advantage, however, may not arise from the specific structure of the embedded motifs, but from the mere presence of a strongly coupled local skeleton within an otherwise weakly coupled random network, irrespective of how those local connections are arranged. To distinguish between these possibilities, we perform motif rewiring. For each motif class, the intra-motif synapses are randomly rearranged while preserving their number and weight, thereby removing the motif-specific organization without changing the total strength of local coupling. The inset of Fig.~\ref{fig:8}A illustrates this procedure for the M2 subnetwork, with intra-motif synapses shown as yellow edges and inter-motif synapses as dashed grey edges.

Figs.~\ref{fig:8}A,B compare the excitatory-population SNR of the intact ERM and SFM networks against that after rewiring individual motif classes, or all simultaneously; the intact networks serve as baselines. To isolate the effect of motif arrangement, we use a smaller network ($N=500$) and a coupling regime in which the motif skeleton contributes more strongly to network dynamics ($w_\mathrm{intra}=1.75$, $w_\mathrm{inter}=0.5$, and $p=0.02$).

In both the ERM and SFM excitatory populations, rewiring M2 gives the largest SNR reduction: $\mathrm{SNR}^*$ falls from $\approx 24$ to $\approx 18$ in ERM and from $\approx 40$ to $\approx 31$ in SFM. Rewiring M3c is next, lowering $\mathrm{SNR}^*$ to $\approx 21$ (ERM) and $\approx 35$ (SFM); rewiring M3b gives only a modest decrease. Rewiring all classes simultaneously reduces $\langle\mathrm{SNR}\rangle_D$ relative to intact, significantly in both networks ($p<0.025$). By contrast, rewiring M3a or M4 increases SNR in both networks: in ERM the increase is larger after M4 than M3a rewiring, whereas in SFM the two are comparable.

\subsection{Signal transmission}
\label{sec:information_transmission}

We next examine how network structure influences the efficacy of stimulus-driven signal transmission. In each trial, 20 excitatory neurons are selected either at random from the full excitatory population or from a specific motif class, and driven by 300 independent Poisson afferents. The afferent rate is $\lambda(t)=25~\mathrm{Hz}$ during stimulation and zero otherwise. Each Poisson spike generates an exponentially decaying current pulse, giving the total stimulus current
$$
I_{\mathrm{poiss}}(t)
=
\sum_{t_s\in\mathcal{S}}
0.03
\exp\!\left(-\frac{t-t_s}{6~\mathrm{ms}}\right)
\mathcal{H}(t-t_s),
$$
where $\mathcal{S}$ is the set of Poisson spike times and $\mathcal{H}$ is the Heaviside function; $I_{\mathrm{poiss}}(t)$ is added to Eq.~\eqref{v} for stimulated neurons. The stimulus is applied during two $25~\mathrm{ms}$ epochs, $300\text{-}325~\mathrm{ms}$ and $700\text{-}725~\mathrm{ms}$, within a $1000~\mathrm{ms}$ simulation. The epochs are kept brief to elicit a transient response while avoiding runaway network activation.

\begin{figure}[t]
\centering
\includegraphics[width=1\linewidth]{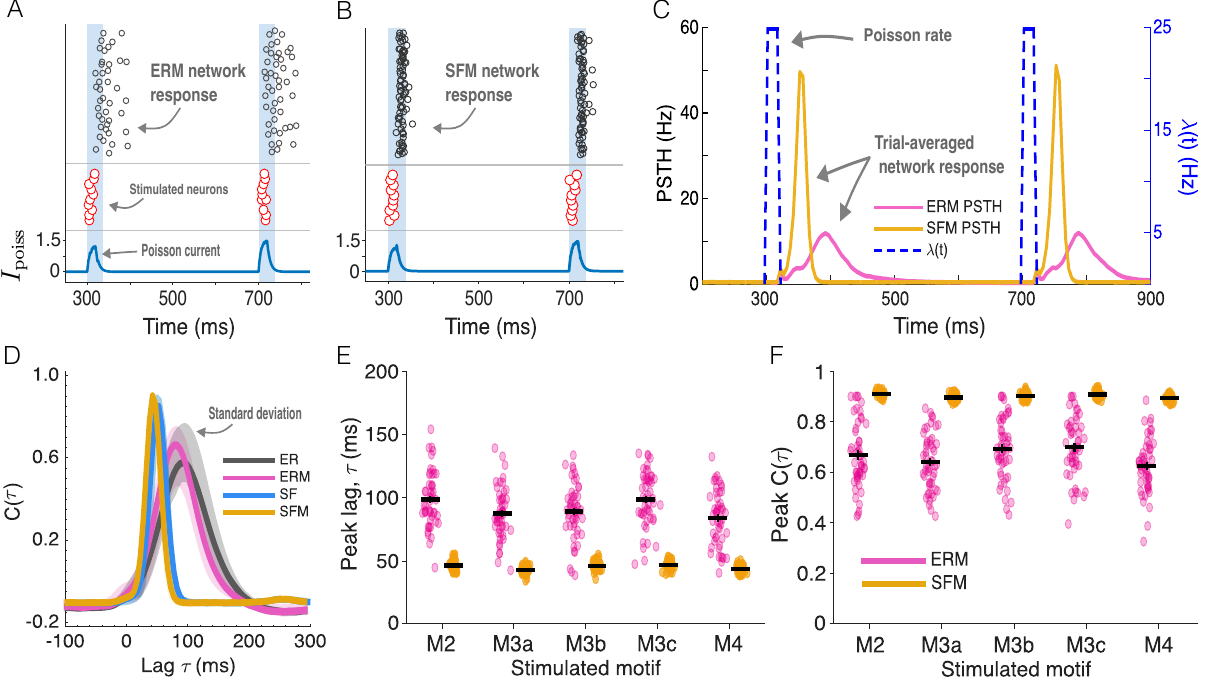}
\caption{Information transmission under Poisson stimulation ($p=0.02$, $w_\mathrm{intra}=1$, $w_\mathrm{inter}=0.25$). (A,B) Raster responses of ERM (A) and SFM (B): Poisson current $I_{\mathrm{poiss}}$ (blue), stimulated neurons (red), network response (black). (C) Poisson rate $\lambda(t)$ and trial-averaged PSTH for ERM and SFM. (D) Trial-averaged $C(\tau)$ between $\lambda(t)$ and PSTH (shaded: trial-to-trial standard deviation). (E,F) Peak lag and peak $C(\tau)$ across stimulated motif classes.}
\label{fig:9}
\end{figure}
Representative responses are shown in Figs.~\ref{fig:9}A,B for the ERM (A) and SFM (B) networks. Each panel displays the Poisson input current $I_{\mathrm{poiss}}(t)$, with red raster points denoting spikes from the stimulated neurons, black points the rest of the network. The population-level responses are summarized in Fig.~\ref{fig:9}C, which shows the on- and off-epochs of the afferent rate $\lambda(t)$ (dotted line) together with the trial-averaged population PSTHs of the ERM (pink) and SFM (orange) networks.

The ERM network produces a broad, temporally dispersed response, its PSTH peaking $\approx 100~\mathrm{ms}$ after stimulus onset. The SFM network, by contrast, produces a sharper, larger peak at $\approx 50~\mathrm{ms}$, reaching $\approx 47~\mathrm{Hz}$ against $\approx 12~\mathrm{Hz}$ for ERM (Fig.~\ref{fig:9}A-C).

Following \cite{vogels2005signal}, we quantify the similarity between the external input and network response using their normalized cross-correlation. For each trial, the correlation between the Poisson rate signal, $\lambda(t)$, and the population response rate, $r(t)$, is defined as
\begin{equation}
C(\tau)=
\frac{
\left\langle
\left(\lambda(t)-\bar{\lambda}\right)
\left(r(t+\tau)-\bar{r}\right)
\right\rangle_t
}{
\sqrt{
\left\langle\left(\lambda(t)-\bar{\lambda}\right)^2\right\rangle_t
\left\langle\left(r(t+\tau)-\bar{r}\right)^2\right\rangle_t
}
},
\end{equation} 
where $\langle\cdot\rangle_t$ denotes a temporal average, and $\bar{\lambda}$ and $\bar{r}$ are the temporal means of the Poisson rate signal and population response, respectively. The cross-correlation curves are then averaged across trials and shown in Fig.~\ref{fig:9}D, with shaded regions denoting the trial-to-trial standard deviation.

The ERM network produces a broad maximum of $C(\tau)\approx 0.6$ at $\tau\approx 90\text{-}100~\mathrm{ms}$, whereas the SFM network produces a sharper, larger one of $C(\tau)\approx 0.9$ at $\tau\approx 40\text{-}50~\mathrm{ms}$. The SFM response is therefore more strongly correlated with the input and concentrated at a shorter, more precisely defined lag. The broader peak and larger variability of the ERM curve indicate a slower, less temporally consistent response across trials.

Compared with the effect of background topology, motif embedding has a modest influence on signal transmission. The SF and SFM networks produce similar $C(\tau)$ profiles, the SFM curve reaching a slightly earlier and higher maximum---consistent with a small improvement in input-response correlation and response latency. The ER and ERM networks show the same trend: the ER peak occurs $\approx 15~\mathrm{ms}$ later and $\approx 0.05$ lower than the ERM peak. These differences are small relative to the trial-to-trial variability, however, and should be read as trends unless statistical comparison of the peak amplitudes and lags confirms their significance.

We next examine motif-specific signal transmission. As shown in Figs.~\ref{fig:9}E,F, stimulation of every motif class produces a shorter peak lag and a larger maximum of $C(\tau)$ in the SFM network (orange points) than in the ERM network (pink points). The SFM measurements also show substantially less trial-to-trial variability, indicating more reliable signal transmission.

Differences among individual motif classes are less clear. In the ERM network, stimulation of M3b and M3c gives slightly larger peak correlations than of M2, M3a, or M4, but the variability is too large to establish a robust motif-specific ordering. Taken together, these results show that the SF background supports faster and more reliable signal transmission than the ER background, while motif embedding provides an additional but modest benefit, more pronounced in the ER-based than the SF-based topology.

\subsection{Scale--free ablation}
\label{sec:ablation}
Having established that the SF/SFM networks support more effective signal transmission than their ER/ERM counterparts, we examine the extent to which this depends on the hub-centred organization of SF topology. To this end, we compare the SFM network with hub-ablated, node-ablated, and edge-ablated variants.

Hub ablation removes the top $7.5\%$ of excitatory neurons ranked by total degree, along with their incident synapses. Node ablation removes the same number of excitatory neurons and their incident synapses by selecting the neurons at random. Edge ablation removes the same number of synapses as hub ablation, selected at random from the intact network. Node ablation thereby controls for the number of neurons removed, edge ablation for the number of synapses.

\begin{figure}[th]
\centering
\includegraphics[width=1\linewidth]{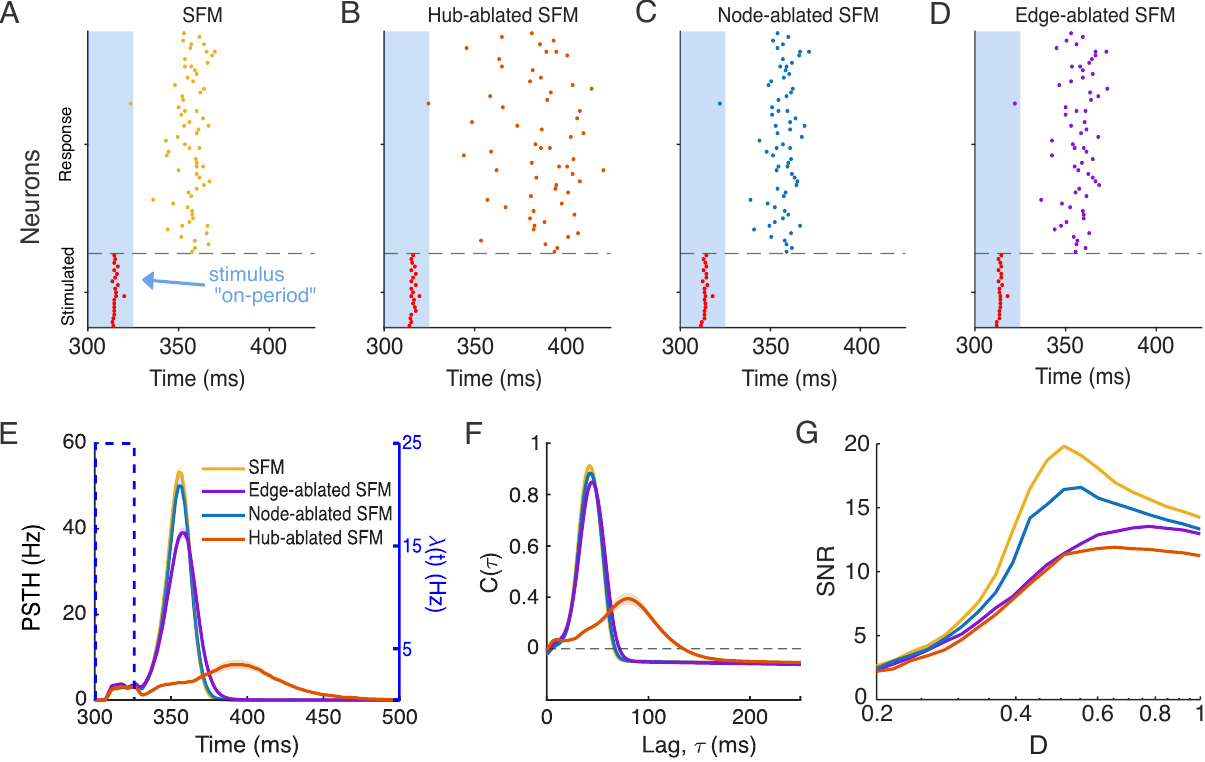}
\caption{Effects of ablation on signal transmission and coherence ($p=0.02$, $w_\mathrm{intra}=1$, $w_\mathrm{inter}=0.25$). (A--D) Raster responses for SFM (A), hub--ablated (B), node--ablated (C), and edge--ablated (D) SFM; shaded region: stimulus on--period; red: stimulated neurons; black: network response. (E) $\lambda(t)$ and response PSTHs. (F) $C(\tau)$ between $\lambda(t)$ and PSTH. (G) SNR vs.\ $D$. Colours: intact (yellow), edge--ablated (purple), node--ablated (blue), hub--ablated (red).}
\label{fig:10}
\end{figure}

We apply the stimulation protocol of Sec.~\ref{sec:information_transmission} to each network. Representative raster plots are shown in Fig.~\ref{fig:10}A--D for the intact SFM (A), hub-ablated SFM (B), node-ablated SFM (C), and edge-ablated SFM (D) networks. Hub ablation produces the clearest qualitative change: the intact, node-ablated, and edge-ablated networks respond synchronously and $\approx25 \ \mathrm{ms}$ after the stimulus onset, whereas the hub-ablated network produces a more dispersed, delayed, and less synchronized response. 

The population PSTHs and $C(\tau)$ curves reflect this (Fig.~\ref{fig:10}E,F). The intact SFM network has the largest PSTH peak, reaching $\approx 50~\mathrm{Hz}$ at $t\approx 350~\mathrm{ms}$ (50~ms after onset). The node-ablated network peaks at a similar latency but slightly lower; the edge-ablated network reaches $\approx 38~\mathrm{Hz}$ with a slightly longer delay. Hub ablation is the most damaging: the PSTH reaches only $\approx 10~\mathrm{Hz}$, peaking at $t\approx 400~\mathrm{ms}$ (100~ms after onset). The $C(\tau)$ curves follow the same pattern (Fig.~\ref{fig:10}F): the intact network has the largest, sharpest peak, followed by the node- and edge-ablated networks, with the hub-ablated network lower, broader, and more delayed.

Returning to network coherence, we examine the SNR (Eq.~\ref{SNR}) of each network as a function of $D$ (Fig.~\ref{fig:10}G). The intact SFM network has the highest peak, $\mathrm{SNR}^*\approx 20$ at $D^*\approx 0.5$. The node-ablated network has a similar $D^*$ but a lower peak, $\mathrm{SNR}^*\approx 17$. Edge and hub ablation produce the largest reductions: the edge-ablated network reaches $\mathrm{SNR}^*\approx 13$ at $D^*\approx 0.7$, and the hub-ablated network only $\mathrm{SNR}^*\approx 11$ at $D^*\approx 0.6$.

The greater impairment under hub ablation than under either control indicates that this effect cannot be explained by the number of neurons or synapses removed alone. The node comparison establishes that removing highly connected neurons is more disruptive than removing the same number at random; the edge comparison, that the effect persists even when the synapse count is matched. The impairment therefore reflects the hub-centred organization of the SFM network: concentrating connections in a few hubs makes their removal disproportionately damaging to signal transmission and coherence.

\section{Discussion}
\label{sec:discussion}
The central question of this study was how local motif organization and global network topology jointly shape noise-driven population activity. All four architectures exhibit coherence resonance, but both its magnitude and optimal noise intensity $D^*$ depend on network structure. Motif embedding increases coherence and lowers $D^*$ in both ER- and SF-based networks. SF and SFM nevertheless attain substantially greater absolute coherence than ER and ERM, and the SF background supports earlier, larger, and more temporally concentrated stimulus responses. Thus, local motif structure provides a measurable contribution in both backgrounds, while global degree heterogeneity produces the larger separation in population-level coherence and transmission under the conditions examined here.

The stronger coherence of the SF-based networks is consistent with previous work showing that heterogeneous connectivity can support coherent neural activity more effectively than homogeneous random topology~\cite{Tonjes2021CoherenceResonance,grinstein2005synchronous,mcgraw2003topology}. Motif embedding gives a smaller but statistically supported increase in noise-averaged SNR ($\langle\mathrm{SNR}\rangle_D$) and lowers $D^*$ in both topologies. At the peak, motif embedding raises $\mathrm{SNR}^*$ by $\approx 10.7\%$ from ER to ERM, compared with $\approx 2.9\%$ from SF to SFM. The larger proportional gain in the ER-based network is consistent with strong local organization partly compensating for a background that otherwise supports less coherent activity; in the SF-based network, the background already produces high coherence, leaving a smaller marginal effect from the same motif skeleton.

The coupling experiments show that these local and global contributions depend on dynamical regime. At weak noise or weak intra-motif coupling, motifs in SFM are more coherent than the same motifs in ERM, indicating that the SF background supports local motif activity when internal coupling alone is insufficient. As intra-motif coupling increases, the motif-level responses converge even while the network-level SNR remains separated. Likewise, at fixed total coupling strength, the SFM advantage is largest in the inter-motif-dominated regime and decreases as coupling shifts toward intra-motif connections. Strong local coupling can therefore organize coherent activity within individual motifs in either background, while population-level coordination remains dependent on global topology. Connection density and coupling strength also affect this topology gap differently: increasing density makes the networks more similar, whereas increasing coupling strength amplifies differences already present in their wiring.

The rewiring experiments show that a strongly coupled local skeleton is not sufficient, by itself, to increase coherence. Rewiring M2 and M3c while preserving the number and weight of their intra-motif synapses reduces network coherence, whereas rewiring M3a and M4 increases it. Rewiring all motif classes simultaneously produces a significant net reduction in $\langle\mathrm{SNR}\rangle_D$. The effect of motif embedding therefore depends partly on the paths formed by the local connections, not only on their number or strength. The opposing effects across motif classes also show that there is no simple division between motifs that uniformly promote and motifs that uniformly suppress coherence. The present rewiring protocol does not isolate which structural features---for example, feedback, path redundancy, or the distribution of incoming and outgoing connections---produce these differences.

The stimulus-driven results place greater weight on global topology than on motif identity. Under brief Poisson stimulation, SFM produces an earlier, larger, and more temporally concentrated population response than ERM, with a larger cross-correlation peak and less trial-to-trial variability. By contrast, the differences between each motif-embedded network and its non-motif control are small relative to the observed variability and should be interpreted as trends unless direct statistical tests of response amplitude and lag establish otherwise. Stimulating different motif classes likewise produces no robust ordering. Under the present protocol, background topology is therefore more strongly associated with transmission efficacy than either motif embedding or the identity of the stimulated motif.

The ablation results indicate that hubs contribute to this topological effect. Removing the highest-degree excitatory neurons weakens and delays the stimulus response and reduces noise-induced coherence more than removing the same number of neurons at random, and more than removing a matched number of synapses at random. The larger impairment under targeted hub removal therefore cannot be explained by the number of neurons or synapses removed alone. It instead reflects, at least in part, the concentration of connections around highly connected neurons. This pattern is consistent with the known robustness of heterogeneous networks to random failure and their vulnerability to targeted disruption~\cite{albert2000error,gal2021role}. The ablation results do not establish that hubs explain the entire SF advantage, but they show that hub integrity makes a contribution not reproduced by the node- or edge-matched controls.

Taken together, the motif- and network-level analyses distinguish two complementary scales of organization. Once intra-motif coupling is sufficiently strong, motif-level coherence becomes similar across the ER and SF backgrounds, even though population-level coherence remains different. Local connection arrangement can therefore determine how coherent activity forms within a motif, while global topology continues to govern how activity is coordinated and propagated across the network. Networks with similar synapse counts and similar local connection strengths need not be dynamically equivalent when those connections form different paths.

Several limitations constrain these conclusions. The ER and preferential-attachment SF networks are idealized topologies, and the motif skeleton was imposed rather than generated through development or activity-dependent plasticity. The Izhikevich neuron and current-based synapses also omit biophysical detail captured by conductance-based models. Future work could address these limitations with more biophysically detailed neuronal and synaptic models, empirically constrained or alternative heterogeneous topologies, and a broader range of network sizes, motif distributions, and hub-ablation fractions. Models in which motifs and hubs emerge through activity-dependent plasticity or network growth would help establish whether the same structural effects arise without being imposed in advance. Within these limitations, the present results support the conclusion that noise-induced coherence depends not only on synapse number and strength, but also on how connections are arranged locally and distributed globally across the network.

\clearpage
\appendix

\section{Simulation parameters}
\label{app:parameters}

\begin{table}[htbp]
\centering
\caption{Baseline parameters used in the simulations.}
\label{tab:params}
\small
\setlength{\tabcolsep}{4pt}
\renewcommand{\arraystretch}{1.1}
\begin{tabular}{llll}
\hline
Parameter & Description & ER / SF & ERM / SFM \\
\hline
$\tau_{\mathrm{inh}}$ & Inhibitory synaptic decay & $6~\mathrm{ms}$ & $6~\mathrm{ms}$ \\
$\tau_{\mathrm{inter}}$ & Inter--motif excitatory decay & $3~\mathrm{ms}$ & $3~\mathrm{ms}$ \\
$\tau_{\mathrm{intra}}$ & Intra--motif excitatory decay & --- & $4~\mathrm{ms}$ \\
$\tau^{(d)}_{\mathrm{inh}}$ & Inhibitory delay & $\mathcal{U}[0.5,2]\,\mathrm{ms}$ & $\mathcal{U}[0.5,2]\,\mathrm{ms}$ \\
$\tau^{(d)}_{\mathrm{inter}}$ & Inter--motif excitatory delay & $\mathcal{U}[1,5]\,\mathrm{ms}$ & $\mathcal{U}[1,5]\,\mathrm{ms}$ \\
$\tau^{(d)}_{\mathrm{intra}}$ & Intra--motif excitatory delay & --- & $0\,\mathrm{ms}$ \\
$\tau_{\mathrm{pre}}$ & Presynaptic STDP trace decay & $20~\mathrm{ms}$ & $20~\mathrm{ms}$ \\
$\tau_{\mathrm{post}}$ & Postsynaptic STDP trace decay & $20~\mathrm{ms}$ & $20~\mathrm{ms}$ \\
$dA_{\mathrm{pre}}$ & Presynaptic trace increment & $0.01$ & $0.01$ \\
$dA_{\mathrm{post}}$ & Postsynaptic trace increment & $-0.012$ & $-0.012$ \\
$W_{\max}$ & Maximum excitatory weight & $10$ & $10$ \\
\hline
\end{tabular}
\end{table}

\newpage

\bibliographystyle{unsrt}
\bibliography{references}

\end{document}